\documentstyle[prb,aps,preprint,epsf]{revtex}
\newcommand{\bea}{\begin{eqnarray}}
\newcommand{\eea}{\end{eqnarray}}
\newcommand{\be}{\begin{equation}}
\newcommand{\ee}{\end{equation}}

\begin{document}

\begin{center}
{\bf {\large \ THERMAL EXCITATIONS OF FRUSTRATED XY SPINS IN TWO DIMENSIONS}}%
\\{\today} \vspace{.5 cm}

{\it M. Benakli}\\ 
\vspace{.5cm}            
                                                                                Department of Physics, Condensed Matter
Section, ICTP, P.O. Box 586, 34014 Trieste, Italy\\                                \vspace{0.5cm}                                                                                                                                                 {\it H. Zheng, M. Gabay}

\vspace{.5 cm}

Laboratoire de Physique des Solides \footnote{%
Laboratoire associ\'e au CNRS}, Universit\'e de Paris-Sud,\\B\^atiment 510,
91405 Orsay Cedex, France                                                       

\end{center}
\begin{abstract}

We present a new variational approach to the study of phase transitions in
frustrated 2D XY models. In the spirit of Villain's approach for the
ferromagnetic case we divide thermal excitations into a low temperature long
wavelength part (LW) and a high temperature short wavelength part (SW). In
the present work we mainly deal with LW excitations and we explicitly
consider the cases of the fully frustrated triangular (FFTXY) and square (
FFSQXY) XY models. The novel aspect of our method is that it preserves the
coupling between phase (spin angles) and chiral degrees of freedom. LW
fluctuations consist of coupled phase and chiral excitations. As a result,
we find that for frustrated systems the effective interactions between phase
variables is long range and oscillatory in contrast to the unfrustrated
problem. Using Monte Carlo (MC) simulations we show that our analytical
calculations produce accurate results at all temperature $T$; this is seen
at low $T$ in the spin wave stiffness constant and in the staggered
chirality; this is also the case near $T_c$: transitions are driven by the
SW part associated with domain walls and vortices, but the coupling between
phase and chiral variables is still relevant in the critical region. In that
regime our analytical results yield the correct $T$ dependence for bare
couplings (given by the LW fluctuations) such as the Coulomb gas temperature 
$T_{CG}$ of the frustrated XY models . In particular we find
that $T_{CG}$ tracks chiral rather than phase fluctuations. Our results
provides support for a single phase transition scenario in the FFTXY and
FFSQXY models.

{\bf PACS Numbers 75.10-b, 75.10.Hk}

\end{abstract}
\newpage

\section*{INTRODUCTION}

\noindent

Frustrated magnetic systems have been extensively studied, in part because
they constitute non-disordered versions of spin-glasses\cite{V1,D1}. They
display rich low-temperature phases and remarkable phase transitions since
frustration modifies the naive symmetry of the Hamiltonian. For spatial
dimensions $D\geq 3$, Kawamura found by renormalization group (RG)
techniques that frustrated $O(n)$ spin models belong to a new, ``chiral''
universality class\cite{Kawa1}. For $D=2$ and XY spins, which is our present
concern, phase transitions are dominated by defects. Frustration results in
additional chiral variables, which generate a discrete symmetry. In the
fully frustrated case the Hamiltonian for a square lattice is believed to
possess an $O(2)\times Z_2$ symmetry (see for instance Ref \cite{YD}); for
the triangular lattice (FFTXY) one has the extra $C_{3V}$ symmetry
associated with the permutation of the three sublattices, thus adding the
possibility of a Potts transition \cite{DHL}. The transition associated with
the $O(2)$ part (phases i.e. angular variables) would be Kosterlitz-Thouless
(K-T)-like at a temperature $T_{KT}$\cite{KT} and the discrete $Z_2$ part
(chiral variables) would be broken below a temperature $T_{DS}$. There is an
ongoing controversy concerning the order in which these transitions should
take place. RG calculations suggest that $T_C=T_{KT}=T_{DS}$ but two
transitions are not ruled out\cite{YD,GKP,LKG,LT,KORSHU} : in the single
phase transition scenario, measurements of critical exponents for the chiral
and of the central charge using MC and MC transfer matrix techniques \cite
{NGK} reveal non Ising behavior, providing support for Kawamura's claim of a
new universality class even in 2D. This is also suggested by studies based
on the selective breaking of certain symmetries\cite{BD,GG,EH,SGZ,ZSGB}.
Some MC studies performed on the FFTXY \cite{SS,DHL,SGZ} and on the fully
frustrated square (FFSQXY) models \cite{BD} yield a single phase transition;
yet other MC simulations for the FFSQXY model and for the $1/2$ integer
Coulomb gas give two phase transitions very close in temperature \cite
{GR,RSJ,JRL,SLKL,O1}.

In view of these unsettled issues the present paper has two objectives :

First we would like to give a quantitative description of the relevant
excitations in FF systems. In doing so, we wish to assess the importance of
the coupling between phase and chiral degrees of freedom at low temperature (%
$T$) and in the critical region. This would allow us to identify the nature
of the critical fluctuations and to decide whether one should expect two
phase transitions or just one.

Second we would like to test if the thermodynamic properties of the FFTXY
and of the FFSQXY models are similar or not, and in particular if the nature
of the phase transitions is different or the same for the two systems.

\noindent

\noindent

In order to get some insight into these issues we present a new analytical
approach to study 2D XY frustrated systems. It is inspired by Villain's
analysis for ferromagnetic (F) systems \cite{V2} where: (a) the partition
and correlation functions are products of a Long Wavelength (LW) -- spin
waves -- and of a Short Wavelength (SW) -- vortex -- contribution \cite{JKKN}%
; and (b) the long wavelength part is mapped {\it quantitatively} (by
perturbation theory or variational scheme) onto the low temperature
contribution of the original cosine Hamiltonian \cite{VPL}. Steps (a) and
(b) allow one to compute accurately all relevant thermodynamic quantities
from $T=0$ up to $T_c$. Our results are best summarized on the figures. In
section I we begin with a brief discussion of steps (a) and (b) for the
unfrustrated case and we show that a simple variational approach -- the self-consistent harmonic approximation (SCHA) \cite{VPL} -- yields
quantitative agreement with MC at all $T$ so long as one considers
thermodynamic variables sensitive to LW excitations (Fig (1)). Extending the
method to FF systems produces incorrect results even at very low $T$ (Fig
(2)). Failure is due to the fact that a naive application of the variational
approach eliminates chiral fluctuations. In section II we set up a new variational method ( we call it NSCHA for new SCHA )
which explicitly preserves the coupling between phase and chiral degrees of
freedom. As a result LW excitations consist of coupled spin waves (phase)
and polar (chiral) fluctuations. This feature causes the effective
interactions between phase variables to be long range and to oscillate in
sign (Figs (3)-(4)). One may contrast this behavior with the unfrustrated case
where phase couplings remain short range and positive in sign. Focusing on
LW fluctuations, in section III we compare our results to MC simulations
performed on the FFTXY and FFSQXY systems. For the FFTXY model we have done
MC calculations for sizes up to $60\times 60$, using $10^5-10^6$ MCS/spin
and fluctuating boundary conditions\cite{SGZ,O2}. For the FFSQXY lattice we
have used available data from the literature : this is justified since we
only use in a quantitative fashion data pertaining to LW excitations (i.e.
results which are common to all studies ). For all $T$ MC simulations and
analytical results agree closely: this is apparent in the LW contributions
to the stiffness constant and to the chiral order parameter (Figs (5)-(8)).
Above a characteristic temperature $T^{*}$ defects become important and they
ultimately drive the transition: we introduce a variable $\tau $ equ (\ref
{AMPCHI}) which allows to track chiral domain walls. Fig (7) shows that they
become relevant above $T^{*}$. Furthermore Figs (5) and (6) show that for $%
T>T^{*}$ domains affect both the chiral order parameter and the spinwave
stiffness constant. The relevance of the coupling between phase and chiral
variables near the transitions is also visible on Fig (8) where we can see
the agreement between analytical and MC predictions for the bare couplings
-- here the Coulomb gas temperature $T_{CG}$ (equ \ref{JCG})--. Noteworthy
is the fact that $T_{CG}$ is connected to chiral variables in FF systems
whereas it is a bare coupling constant for phase fluctuations in the
unfrustrated situation (Fig (1)); this point is important in view of the fact
that MC studies of FF systems assume that $T_{CG}$ is a bare coupling for
the {\it phase} variables (see discussion in section III). Our study thus
suggests that the vanishing of the spinwave stiffness and of the chiral
order parameter occur at the same T, for the FFTXY and for the FFSQXY models.

\section{ the Self Consistent Harmonic Approximation}

\subsection{ The ferromagnetic case}

Using standard notation the Hamiltonian reads

\begin{equation}  \label{H}
{\cal H} = -\sum_{<i,j>}J_{ij}cos(\theta_i-\theta_j)
\end{equation}



In the ferromagnetic case ($J_{ij} =J>0$ for nearest neighbor pairs) Villain
replaces the cosine potential by a parabolic form; 
$e^{\beta Jcos(\theta _i-\theta _j)}$
is approximated by :

\begin{equation}
\label{VP} 
Const. \sum_{n_{ij}} e^{-\beta J_V(\theta_i-\theta_j -2\pi n_{ij})^2 }
\end{equation}

where the integers $n_{ij}$ express the periodicity of the original
interaction. The main features of the ''Villain form '' are that it includes
both LW excitations (spin waves connected to the phases $\theta $) and SW
excitations (vortices connected to the lattice curl of the $n$) and that the
partition function is the product of the LW and SW parts. Below $T_{KT}$ the
vortex part is essentially irrelevant ( it simply introduces a dielectric
constant $\epsilon _V\sim 1+e^{-J/T}$) and LW properties are described by
an harmonic spin wave hamiltonian with a spinwave stiffness constant $%
\gamma=J_V/\epsilon _V$. The importance of Villain's form stems from the
fact that by applying the Migdal-Kadanoff scheme to the original cosine
interaction, one iterates towards an effective harmonic spin wave theory
below $T_{KT}$ \cite{JKKN}. The shortcoming of (2) is that the coupling
constant of the LW part is temperature independent whereas the original
cosine form introduces interactions between spin waves. To deal with this
issue one may use the self consistent harmonic approximation (SCHA)\cite{VPL}
: for the LW part one uses a variational hamiltonian

\begin{equation}  \label{HVAR}
{\cal H}_0={\frac{1}{2}}\sum_{<i,j>}\widetilde J_{ij}(\theta_i -\theta_j)^2
\end{equation}

Anharmonicites of the cosine potential translate into a temperature
dependence of $\widetilde J_{ij}$.

The variational free energy is given by $F_{var}=F_0 +<{\cal H} -{\cal H}_0>_{{\cal H}_0}$ ($F_0$ is the free energy for hamiltonian ${\cal H}_0$ ) and reads

\begin{equation}
F_{var}=-\frac 12\sum_{<i,j>}{\widetilde{J}_{ij}}y_{ij}-%
\sum_{<i,j>}J_{ij}e^{-{\frac 12}y_{ij}}+\frac T2\log Det({\widetilde{{\cal J}%
}}/T)  \label{FVAR}
\end{equation}
${\widetilde{{\cal J}}}$ is the matrix with diagonal elements $\sum_k\widetilde{J}_{ik}$
and off diagonal elements $-\widetilde{J}_{ij}$.
The quantities $y_{ij}=<(\theta _i-\theta _j)^2>_{{\cal H}_0}$ are
themselves functions of the variational parameters $\widetilde{J}_{ij}$.
Therefore we may use $y_{ij}$ as alternative variational parameters. The
variational equations 
read :

\begin{equation}
\widetilde{J}_{ij}=J_{ij}e^{-{\frac 12}y_{ij}}  \label{JSCHA}
\end{equation}
with

\begin{equation}
y_{ij}={\frac T{2\pi ^2}}\int\!\!\!\int_{BZ}
d_2q{\frac{(1-cos\vec q.(\vec r_i-\vec r_j))}{%
\widetilde{J}(0)-\widetilde{J}(\vec q)}}  \label{CORREL}
\end{equation}

Here $\widetilde J(\vec q)$ is the Fourier transform of $\widetilde J_{ij}$.
In the ferromagnetic case, the effective interactions $\widetilde J_{ij}$
only couple nearest neighbors. Their magnitude $\widetilde J(T)$ is given by

\begin{equation}
\widetilde{J}(T)=Je^{-{\frac T{z\widetilde{J}(T)}}}  \label{JTILDESCHA}
\end{equation}
($z$ is the number of nearest neighbors of the lattice). One may then
compute the spinwave stiffness matrix; its elements are given by the second
derivatives of the free energy with respect to uniform twists of the phase 
\cite{OJ}. For isotropic lattices the matrix is diagonal and all the
elements are equal. In SCHA the constant is $\gamma _{SCHA}=\widetilde{J}(T)$%
. We may then simply replace $J_V$ in equ (\ref{VP}) above by $\gamma
_{SCHA} $. Figs (1a) and (1b) show the temperature dependence of the SCHA
stiffness for the square (SQ) and triangular (TR) lattices along with the MC
result (denoted by $\gamma (T)$). At low T vortices contribute with a
probability of order $e^{-zJ/T}$ so that SCHA and MC agree quite well. Near $%
T_{KT}\approx 0.892J$ (SQ lattice \cite{WM,SM,O3}) or $1.446J$ (TR lattice 
\cite{BC}) vortices cause a drop in $\gamma (T)$. Since SCHA only describes
LW fluctuations it fails to produce a fall-off. \vskip 1cm \noindent

\subsection{ the Fully Frustrated case}

The previous analysis is easily extended to the situation where spins are
non collinear in equilibrium \cite{ABV}. We rewrite the $\theta_i$ of equ (\ref{H}) as:

\begin{equation}
\theta _i={\theta _i}^0+\varphi _i  \label{ANGLES}
\end{equation}
where ${\theta _i}^0=<\theta _i>_{{\cal H}_0}$ and the variational
hamiltonian is 
\begin{equation}
{\cal H}_0={\frac 12}\sum_{<i,j>}\widetilde{J}_{ij}(\varphi _i-\varphi _j)^2
\label{HPHI}
\end{equation}

In addition to the parameters $y_{ij}=<(\varphi _i-\varphi _j)^2>_{{\cal H}%
_0}$ one has the extra variables ${\theta _i}^0$. The variational equations
now read 
\begin{equation}
\widetilde{J}_{ij}=J_{ij}cos({\theta _i}^0-{\theta _j}^0).e^{-{\frac 12}%
y_{ij}}  \label{JFRUSSCHA}
\end{equation}

\begin{equation}
\sum_{<i,j>}\widetilde{J}_{ij}tan({\theta _i}^0-{\theta _j}^0)=0
\label{TETASCHA}
\end{equation}
and $y_{ij}$ is given by equ.(\ref{CORREL}). Denoting by $\vec r_i=(x_i,y_i)$
the vector connecting the origin of the lattice to site $i$ and by $\vec u%
_{ij}$ the vector connecting nearest neighbor sites $i$ and $j$, the solution
${\theta _i}^0$ to these equations is independent of $T$ and given by:

\begin{equation}
{\theta _j}^0-{\theta _i}^0\equiv {\theta }^0(\vec r_i+\vec u_{ij})-{\theta }%
^0(\vec r_i)=A_{ij}+(-1)^{x_i+y_i}\alpha _{SQ}\;(mod\;2\pi )  \label{ANGSQ}
\end{equation}
for the square lattice, along with the symmetry property

\begin{equation}
{\theta }^0(\vec r_i+\vec u_{ij})-{\theta }^0(\vec r_i)=+({\theta }^0(\vec r%
_i-\vec u_{ij})-{\theta }^0(\vec r_i))  \label{SYMSQ}
\end{equation}
\hskip 3.3cm and by:

\begin{equation}  \label{ANGTR}
{\theta_j}^0-{\theta_i}^0\equiv {\theta}^0(\vec r_i+\vec u_{ij})- {\theta}^0(%
\vec r_i) =\vec Q.\vec u_{ij}= A_{ij}\pm \alpha_{TR}\; (mod \; 2\pi)
\end{equation}
for the triangular lattice, along with the symmetry property

\begin{equation}  \label{SYMTR}
{\theta}^0(\vec r_i +\vec u_{ij}) -{\theta}^0(\vec r_i) =-({\theta}^0(\vec r%
_i -\vec u_{ij})-{\theta}^0(\vec r_i))
\end{equation}

Here $A_{ij}=-A_{ji}=\pi $ if $J_{ij}<0$ and $A_{ij}=0$ if $J_{ij}>0$. $\alpha $
is a lattice dependent, temperature independent quantity. For the FFTXY
lattice ($J_{ij}=-J<0$ ) one has $\vec Q\propto ({\frac{2\pi }3},{\frac{2\pi 
}{\sqrt{3}}})$ so that $\alpha _{TR}={\frac \pi 3}$ . For the FFSQXY lattice
-- the so-called Villain odd model -- ($J_{ij}=-J<0$ every other row along
say the horizontal direction and $J_{ij}=+J>0$ otherwise) one finds $\alpha
_{SQ}={\frac \pi 4}$. In addition $\widetilde{J}_{ij}$ is a {\it nearest
neighbor} interaction of magnitude

\begin{equation}
\widetilde{J}(T)=Jcos(\alpha ).e^{-{\frac T{z\widetilde{J}(T)}}}
\label{JTSCHA}
\end{equation}

Fig (2) shows the SCHA stiffnesses for the FFTXY lattice together with the
MC result. Even at low $T$, $\widetilde J(T) $ and $\gamma(T)$ differ
significantly. The same effect is observed for the FFSQXY lattice. The
reason for this discrepancy is clear : inserting equ (\ref{ANGLES}) into equ
(\ref{H}) gives

\begin{equation}  \label{HFER}
{\cal H} = -\sum_{<i,j>}J_{ij}(cos({\theta_i}^0-{\theta_j}%
^0)cos(\varphi_i-\varphi_j)- sin({\theta_i}^0-{\theta_j}^0)sin(\varphi_i-%
\varphi_j))
\end{equation}

Within SCHA the sin() term of equ (\ref{HFER}) averages to zero. But this
term precisely discriminates the $+\alpha$ solution from the $-\alpha$
solution in equs (\ref{ANGSQ},\ref{ANGTR}) and these two solutions
correspond to the two chiral groundstates of the FF system. As a result SCHA
washes out chiral fluctuations and maps the hamiltonian onto an effective
ferromagnetic phase problem since the $J_{ij}$  are simply renormalized to           $J_{ij}cos(%
{\theta_i}^0-{\theta_j}^0)$.

Our analytical method was thus required to preserve the coupling between the
two chiral states and to allow fluctuations of the chiralities. Next section
shows that our approach then yields accurate results for the FF case.
Furthermore it also improves on the standard SCHA in the unfrustrated case. 
\vskip
2cm \noindent

\section{ NSCHA}

Using equ (\ref{HFER}) the partition function reads :

\begin{equation}
{\cal Z}=Tr_{{\varphi }_i}I_2(\varphi _i-\varphi _j)e^{\beta
\sum_{<i,j>}J_{ij}cos({\theta _i}^0-{\theta _j}^0)cos(\varphi _i-\varphi _j)}
\label{Z}
\end{equation}
where $I_2=e^{-\beta \sum_{<i,j>}J_{ij}sin({\theta _i}^0-{\theta _j}%
^0)sin(\varphi _i-\varphi _j)}$. We rewrite $I_2$ as the sum of a term even
in $\varphi $ plus a term odd in $\varphi $ :

$I_2(x)={\frac{1}{2}}[I_2(x)+I_2(-x)] +{\frac{1}{2}}[I_2(x)-I_2(-x)]$

Now the trace over $\varphi$ in equ (\ref{Z}) is constrained by equ (\ref
{ANGLES}) $\theta_i={\theta_i}^0 + \varphi_i$ but for LW excitations we
expect $\varphi_i$ to fluctuate about $0$, so that we can safely extend the
domain of variation of $\varphi_i$ to the interval $[-\pi,+\pi]$. As a
result the odd term of $I_2$ drops out and the partition function reads

${\cal Z}=Tr_{\varphi _i}e^{-\beta {\cal H}_{eff}}$ where

\begin{equation}
\begin{tabular}{ll}
${\cal H}_{eff}=$ & $-\sum_{<i,j>}J_{ij}cos({\theta _i}^0-{\theta _j}%
^0)cos(\varphi _i-\varphi _j)$ \\ 
& $-TLog[cosh(\sum_{<i,j>}\beta J_{ij}sin({\theta _i}^0-{\theta _j}%
^0)sin(\varphi _i-\varphi _j))]$%
\end{tabular}
\label{HEFF}
\end{equation}

The second term on the r.h.s of equ (\ref{HEFF}) is the new relevant term.
It has the following properties :

${\bf \bullet }$ it is a thermal contribution since it vanishes at $T=0$

${\bf \bullet }$ it is zero for collinear systems (i.e. either unfrustrated
or frustrated but with non chiral configurations)

${\bf \bullet}$ it couples chirality ($\theta^0_i$ -- more precisely $%
J_{ij}sin({\theta_i}^0-{\theta_j}^0)$ for each link (ij) equ (\ref{SIG1})
below -- ) and phase ($\varphi_i$) variables, allowing chiral fluctuations.
Similarly it preserves the symmetry between the two chiral groundstates.

We now compute the variational free energy associated with ${\cal H}_{eff}$
using the trial hamiltonian ${\cal H}_0$ (equ (\ref{HPHI})). To do so we
have to expand the $log(cosh)$ term of ${\cal H}_{eff}$ in power series of
its argument. It is justified since the series amounts to a multipole
expansion as is seen below. Besides, the leading term is the first term,
especially at low T. ${\cal H}_{eff}$ then becomes

\begin{equation}
\begin{tabular}{ll}
${\cal H}_{NSCHA}=$ & $-\sum_{<i,j>}J_{ij}cos({\theta _i}^0-{\theta _j}%
^0)cos(\varphi _i-\varphi _j)$ \\ 
& $-{\frac 1{2T}}\sum_{<i,j>}\sum_{<k,l>}J_{ij}J_{kl}\;sin({\theta _i}^0-{%
\theta _j}^0)sin({\theta _k}^0-{\theta _l}^0)$ \\ 
\multicolumn{1}{c}{} & \multicolumn{1}{c}{$sin(\varphi _i-\varphi
_j)sin(\varphi _k-\varphi _l)$}
\end{tabular}
\end{equation}

The variational equations for -- what we call -- the NSCHA (new SCHA)
ensemble are:

\begin{equation} \label{JNSCHA}
\begin{tabular}{ll}
$\widetilde{J}_{ij}=$ & $J_{ij}\;cos({\theta _i}^0-{\theta _j}^0)\;e^{-{%
\frac 12}y_{ij}}$ \\ 
& $+{\frac 1{2T}}\sum_{k,l}J_{ij}J_{kl}\;sin({\theta _i}^0-{\theta _j}^0)sin(%
{\theta _k}^0-{\theta _l}^0)e^{-{\frac 12}%
(y_{ij}+y_{kl}+y_{ik}+y_{jl}-y_{il}-y_{jk})}$ \\ 
& $+{\frac 1T}\sum_{k,l}J_{ik}J_{jl}\;sin({\theta _i}^0-{\theta _k}^0)sin({%
\theta _j}^0-{\theta _l}^0)$ \\ 
\multicolumn{1}{c}{} & \multicolumn{1}{c}{$cosh(y_{ij}+y_{kl}-y_{il}-y_{jk})%
\;e^{-{\frac 12}(y_{ik}+y_{jl})}$}
\end{tabular}
\end{equation}

\begin{equation}
\begin{tabular}{rr}
$\sum_jJ_{ij}\;sin({\theta _i}^0-{\theta _j}^0)\;e^{-{\frac 12}y_{ij}}$ & 
\\ 
\multicolumn{1}{l}{$-{\frac 1{2T}}\sum_{j,k,l}J_{ij}J_{kl}\;cos({\theta _i}%
^0-{\theta _j}^0)sin({\theta _k}^0-{\theta _l}^0)e^{-{\frac 12}%
(y_{ij}+y_{kl}+y_{ik}+y_{jl}-y_{il}-y_{jk})}$} & \multicolumn{1}{l}{$=0$}
\end{tabular}
\end{equation}

Again $y_{ij}=< (\varphi_i -\varphi_j)^2 >_{{\cal H}_0}$ and

\begin{equation}  \label{CORRELNSCHA}
y_{ij}= {\frac{T}{(2\pi^2)}} \int\!\!\!\int_{BZ}d_2q{\frac{(1-cos\vec q.(%
\vec r_i -\vec r_j))}{\widetilde J(0) - \widetilde J(\vec q)}}
\end{equation}

For the FFTXY and FFTSQXY lattices it is easy to check that $\theta_i^0$ is
a temperature independent quantity and that its value is still given by equs
(\ref{ANGSQ},\ref{ANGTR})

Futhermore equ (\ref{JNSCHA}) shows that $\widetilde J_{ij}$ is no longer a
short range interaction. In fact we find that for all T,

\begin{equation}
\widetilde{J}_{ij}\sim {\frac 1{|\vec r_i-\vec r_j|^z}}  \label{LRA}
\end{equation}
for large distances $r=|\vec r_i-\vec r_j|$ (Fig. 3). This comes about
because $y_{ij}\sim log(r)$ at large distances so that $%
y_{ik}+y_{jl}-y_{il}-y_{jk}\sim {\frac 1{r^2}}$ in equ (\ref{JNSCHA}) : this
contribution is quadrupolar like. Similarly, expanding the log(cosh) term to
next order would produce a higher order multipolar contribution (see also
Appendix A).

In addition the sign of $\widetilde J_{ij}$ varies with the relative
orientation of $i$ and $j$ and, in the case of the FFSQXY lattice, with the
distance between $i$ and $j$ Fig (4a) and (4b).

These features are to be contrasted with the results of SCHA yielding a
positive nearest neighbor $\widetilde J_{ij}$. The coupling between phase
and chiral degrees of freedom has produced an oscillating, ``long range"
interaction between the phase variables. Because of these properties it is
clear that renormalization group analyses (e.g Migdal-Kadanoff) are not
straighforward for FF systems.

Within NSCHA we can compute the phase stiffness constant. Owing to the
isotropy of the lattices we have :

\begin{equation}
\Gamma (T)=\lim_{q_x\to 0}{\frac{(\widetilde{J}(0)-\widetilde{J}(\vec q.\vec 
u_x))}{q_x^2}}  \label{GAMMA}
\end{equation}
where $\vec u_x$ is the unit vector along the horizontal direction of the
lattice. Besides, within this new variational ensemble we also get a
stiffness associated with the canting of the spins; considering a small         variation of the nearest neighbor angle difference                              ${\theta _i}^0-{\theta _j}^0$ from its
equilibrium value in the form $\Delta\vec u_x.\vec u_{ij}$ we get:

\begin{equation} \label{GAMNSCHA}
\begin{tabular}{ll}
$\gamma _{NSCHA}(T)$ & $={\frac{\delta ^2F_{var}}{\delta (\Delta
)^2}}\vert_{\Delta\to 0}$ \\ 
& $={\frac 1N}\left( \sum_{<i,j>}J_{ij}\;cos({\theta _i}^0-{\theta _j}^0)(%
\vec u_{ij}.\vec u_x)^2e^{-{\frac 12}y_{ij}}\right. $ \\ 
& $-{\frac 1T}.\sum_{<i,j>}\sum_{<k,l>}J_{ij}J_{kl}\;(\vec u_{ij}.\vec u_x)(%
\vec u_{kl}.\vec u_x)e^{-\frac 12(y_{ij}+y_{kl}+y_{ik}+y_{jl}-y_{il}-y_{jk})}
$ \\ 
& $\left. \left[ cos({\theta _i}^0-{\theta _j}^0)cos({\theta _k}^0-{\theta _l%
}^0)+sin({\theta _i}^0-{\theta _j}^0)sin({\theta _k}^0-{\theta _l}^0)\right]
\right) $%
\end{tabular}
\end{equation}

It is easy to show that $\gamma _{NSCHA}(T)$ is nothing but the average of
the exact spinwave stiffness $\gamma (T)$\cite{DHL} in the ensemble ${\cal H}%
_0$, i.e 
\begin{equation}\label{GAMMC}
\begin{tabular}{ll}
$\gamma _{NSCHA}(T)=$ & ${\frac 1N}<\sum_{<i,j>}J_{ij}\;cos({\theta _i}-{%
\theta _j})\;(\vec u_{ij}.\vec u_x)^2$ \\ 
& $-{\frac 1T}\;\sum_{<i,j>}\sum_{<k,l>}J_{ij}J_{kl}\;(\vec u_{ij}.\vec u_x)(%
\vec u_{kl}.\vec u_x)\;sin({\theta _i}-{\theta _j})sin({\theta _k}-{\theta _l%
})>_{{\cal H}_0}$%
\end{tabular}
\end{equation}

A plot of $\Gamma (T)$ and $\gamma _{NSCHA}(T)$ versus $T$ is shown for the
TR lattice (Fig (5a)) and for the SQ lattice (Fig (5b)). We note that $%
\Gamma (T)$ and $\gamma _{NSCHA}(T)$ coincide at low $T$. This is explicitly
demonstrated in Appendix B. In particular we find that

\begin{equation}  \label{RIGID}
\gamma_{NSCHA}(T)=\gamma_0(1-{\frac{T}{T_{c0}}})
\end{equation}

${\bf \bullet}$ For the triangular lattice $\gamma_0=\sqrt 3/2 J$ and $%
T_{c0}={\frac{1}{ 4/3-{\frac{3\sqrt 3}{2\pi}}}}J\sim 1.975J$

${\bf \bullet}$ For the square lattice $\gamma_0=\sqrt 2/2J$ and $T_{c0}= {%
\frac{1}{{\frac{\sqrt 2}{2}}-{\frac{\sqrt 2}{2\pi}}}}J\sim 2.075J$

For both cases $T_{c0}\simeq 2J$ (Ref \cite{Min85}).

\vskip 1cm \noindent Another quantity of interest is the staggered chirality

\begin{equation}
\sigma ={\frac 1{{\cal N}_P}}\sum_P{\frac{<\sum_{(k,l)\in P}\sigma _{kl}>_{%
{\cal H}_0}}{\sum_{(k,l)\in P}\sigma _{kl}(T=0)}}  \label{STAGCHI}
\end{equation}
$P$ denotes plaquettes of the same sublattices i.e plaquettes in the same
chiral state at $T=0$.  
The summation $\sum_{(k,l)\in P}$ is performed over the links of plaquette P oriented clockwise.
 $\sigma _{kl}$ is
defined as :

\begin{equation}
\sigma _{kl}=J_{kl}sin(\theta _k-\theta _l)  \label{SIG1}
\end{equation}
(see below for a discussion on the definition of $\sigma _{kl}$ ). 

Using equ (\ref{ANGLES}), we have \\
$\sigma_{kl}=J_{kl}(cos({\theta_k}^0-{\theta_l}%
^0)sin(\varphi_k-\varphi_l)+ sin({\theta_k}^0-{\theta_l}^0)cos(\varphi_k-%
\varphi_l))$ \\ Within NSCHA the $sin(\varphi_k-\varphi_l)$ term drops out and
$\sigma _{kl}(T=0)=J_{kl}(cos({\theta_k}^0-{\theta_l}^0)$ so that

\begin{equation}
\sigma_{NSCHA} =e^{-{\frac 12}y_{kl}}  \label{SIGNSCHA}
\end{equation}
where $k$ and $l$ are {\it nearest neighbors} ($y_{kl}$ has the same value for all the nearest neighbor sites $l$ of any given site $k$). $\sigma_{NSCHA}$      versus $T$ is
plotted on Fig (6) for the TR and SQ lattices.

To summarize the results of this section we see that LW thermal excitations
in fully frustrated lattices are characterized by a strong coupling between
chiral and phase degrees of freedom. The effective interaction between phase
variables is long range and oscillatory -- in contradistinction with the
unfrustrated case --.

Let us now compare our results to those coming from Monte Carlo simulations.

\vskip 2cm \noindent 

\section{ Monte Carlo versus NSCHA}

In order to test the predictions of NSCHA we used the results of Monte Carlo
simulations. For the FFSQXY lattice we took data from the literature insofar
as we did not seek to extract information about critical fluctuations. In
that case there is agreement among the various studies. For the FFTXY and
for the ferromagnetic triangular lattice recent data is rather scarce (Refs 
\cite{SS,DHL,SGZ}) so that we performed our own simulations. We considered
typical lattice sizes of $48\times 48$ and ran $10^5 -10^6$ MCS/spin. In
order to minimize boundary effects we used fluctuating boundary conditions
(Refs \cite{SGZ,O2}). We monitored the following quantities~:

${\bf \bullet}$ the spinwave stiffness constant $\gamma (T)$

${\bf \bullet}$ the staggered chirality $\sigma$ equ (\ref{STAGCHI}):

\noindent several definitions of $\sigma$ have been used in the literature.
One of them is the definition we use here (see also Olsson \cite{O2}),
others are\cite{RSJ}

\begin{equation}
\sigma ^1={\frac 1{{\cal N}_P}}<\sum_PSign(\sum_{(k,l)\in P}\sigma _{kl})>
\label{SIG2}
\end{equation}
where $\sigma _{kl}$ is defined in equ (\ref{SIG1}) and ${{\cal N}_P}$ the
number of plaquettes of each sublattice, or

\begin{equation}
\sigma ^{\prime }={\frac 1{{\cal N}_P}}<\sum_P\sum_{(k,l)\in P}\sigma
_{kl}^{\prime }>  \label{SIGP}
\end{equation}
with \cite{SLKL}

\begin{equation}
\sigma _{kl}^{\prime }={\frac {1}{2\pi} }(\theta _k-\theta _l-A_{kl})
\label{SIGP1}
\end{equation}
or with

\begin{equation}  \label{SIGP2}
\sigma_{kl}^{\prime}={\frac{1}{2\pi}}(\theta_k-\theta_l)
\end{equation}

In equs (\ref{SIGP1},\ref{SIGP2}) the angular determination of the terms in
parenthesis is taken in the interval $\rbrack -\pi,+\pi\rbrack$. All these
definitions lead to the same T dependence for $\sigma$ in the critical
region. For the square lattice this is reported by \cite{RSJ} for instance
and for the triangular lattice this is seen on Fig (7) using the definitions
equs (\ref{SIG1}) and (\ref{SIGP2}).

${\bf \bullet}$ the chirality amplitude $\tau$ :

\begin{equation}
\tau ={\frac 1{{\cal N}_P}}\sum_P{\frac{<Abs(\sum_{(k,l)\in P}\sigma _{kl})>%
}{Abs(\sum_{(k,l)\in P}\sigma _{kl}(T=0))}}  \label{AMPCHI}
\end{equation}
using again the previous definitions for $\sigma _{kl}$. So long as
chiralities are ordered on each sublattice $\tau $ and $\sigma $ coincide.
When domains of the ``wrong'' chiral state form on a given sublattice the
two quantities differ. Thus $\tau $ allows us to track the formation of
domains and domain walls (where the chirality of a plaquette $\sum_{(k,l)\in P}\sigma _{kl}=0$). For the FFTXY lattice for
instance Fig (7) shows that at $T_c$ we have $\sim 30\%$ of positive
chiralities, $\sim 30\%$ of negative chiralities and $\sim 40\%$ of a-chiral
plaquettes on each sublattice.

${\bf \bullet}$ the Coulomb gas temperature $T_{CG}$ :

This quantity monitors the bare (unrenormalized) coupling constant and allows to
define the critical point for the XY model (Refs \cite{Min87,O3}). Within MC
it is given by \cite{O3}

\begin{equation}
T_{CG}={\frac T{{2\pi J_0}}},\;\;\;\;J_0=J<cos(\theta _i-\theta _j)>
\label{JCG}
\end{equation}
for nearest-neighbors i and j.

\vskip 1cm \noindent

Figs (5a) and (5b) show $\Gamma (T),\gamma_{NSCHA}(T)$ and $\gamma (T)$
versus T. The three curves yield the same variation at low $T$. Furthermore,
as could be expected from our previous discussion, $\gamma_{NSCHA}(T)$
tracks $\gamma (T)$ for $T\leq T^*$ ($T^* \sim 0.32J$ for the square lattice
and $T^*\sim 0.35J$ for the triangular lattice); for $T > T^*$ the two
curves move apart. Since NSCHA describes LW excitations but neglects SW
excitations responsible for the transitions, this had to be expected.

Similarly Figs (6) shows a comparison between NSCHA and MC for $\sigma(T)$;
again the agreement is quite good for $T\leq T^*$. Moreover we also see from
Fig (7) that $T^*$ marks the temperature above which domain walls become
important, since $\sigma(T)$ and $\tau(T)$ start to differ for $T\sim T^*$.

At this stage we might worry that the discrepancy between MC and NSCHA
predictions for $T>T^*$ not only marks the point when defects become
important but also signals the breakdown of the variational approach. In
fact NSCHA still yields accurate results for quantities sensitive to LW
fluctuations including in the critical region. We see this by comparing the
MC and the NSCHA $J_0$ entering the definition of the Coulomb gas
temperature. $T_{CG}$ represents the bare (unrenormalized) coupling constant
when LW fluctuations are taken into account.

For instance in the case of the square lattice, Olsson finds that $%
T_{CG}\sim 0.128$\cite{O1} at the KT transition ($T_{KT}\sim 0.446J$); this
value is to be compared with the MC results by Grest \cite{GR} ($T_{CG}\sim
0.126$) and by Lee \cite{JRL} ($T_{CG}\sim 0.1297$) on the half integer
Coulomb gas representation of the FFSQXY.

Fig (8) shows $J_0(T)$ for the square and triangular lattices determined
both in MC and in NSCHA. We notice that:

a) both determinations agree extremely well in the critical regime

b) $J_0$ {\it tracks the chiral variable couplings rather than the phase
variable couplings}~:                                                          

\noindent
Indeed, if we use for $J_0$ the definition given in equ (\ref{JCG}) we find that in the NSCHA ensemble

\begin{equation}  \label{JISING}
{J_0}_{NSCHA}=cos(\theta^0_i-\theta^0_j)\sigma_{NSCHA}
\end{equation}
and $J_0$ is therefore connected to the {\it chiral} variables.
We have seen that the LW contribution to the chirality $\sigma$ -- given by
equ (\ref{SIGNSCHA}) -- does not vanish at the transition ($\sigma$ becomes
zero because of defects) so that ${J_0}_{NSCHA}$ is finite even in the critical regime. For instance 
 for the square lattice NSCHA gives $T_{CG}\sim 0.125$
using $T_{KT}=0.446J$.
By contrast, for the unfrustrated case, equ (\ref{JCG}) gives                   ${J_0}_{SCHA}={\widetilde J}(T)$ (see section I). So $J_0$ is connected to {\it phase} variables then (see also below).

\noindent
In the frustrated case, if we replaced $J_0$ by $\Gamma(T)$ equ (\ref{GAMMA}) we would find too high a value for $T_{CG}$ (namely 0.139) compared to MC.

Similarly, for the triangular lattice MC gives $T_{KT}=0.51J$ (Refs \cite
{DHL,SGZ}) and $T_{CG}~\sim 0.123$ to be compared with the NSCHA prediction     (using equ (\ref{JISING}))                     
$T_{CG} \sim 0.122$. 

This result has direct implications for MC studies : these introduce a
second critical temperature $T_{DS}$ where chiral order vanishes. Some
authors find $T_{KT}> T_{DS} $\cite{RSJ,LT} whereas others predict $T_{KT} <
T_{DS} $ \cite{GR,SLKL,O1}. At $T_{DS}$ critical exponents are found to be
Ising-like by some authors \cite{BD,O1} but non-Ising by others \cite
{RSJ,SLKL}. The magnitude of the jump of the spinwave stiffness constant or
of the dielectric constant seen in MC appears universal for the FFTXY model (%
\cite{Min85}) but non universal for the FFTSQXY model \cite
{Min85,GR,RSJ,JRL,SLKL}. Recently, Olsson has argued that a correct analysis
of the transitions in the case of the FFTSQXY requires extra care due to
their closeness in temperature. As a result he finds a universal jump at $%
T_{KT}$ and similarly Ising exponents at the chiral transition\cite{O1} in
contradistinction with previous authors \cite{Min85,GR,RSJ,JRL,SLKL}. One
should note that the claim of universality or non universality for the KT
transition is based on a scaling a la Minnhagen for the magnitude of the
jump : yet, according to Minnhagen's study this scaling should not hold (one
might even expect a first order transition) given the value of the critical
Coulomb gas temperature corresponding to $T_{KT}$ \cite{Min87,MW}.

Our results show that, because of the coupling between phase and chiral
degrees of freedom, $T_{CG}$ pertains to chiral variables; because of this coupling  one might thus
expect a single phase transition in these systems.

For the unfrustrated case -- e.g in the ferromagnetic limit --, equ (\ref{JCG}) gives                                                            
${J_0}_{SCHA}=\gamma_{SCHA}={\widetilde J}(T)$. Its temperature dependance compares reasonably well with MC (Fig (1)). In fact MC and variational predictions agree extremely well if one compares $J_0$ to $\gamma_{NSCHA}$: 
NSCHA reduces to SCHA for the most part but even in the ferromagnetic case  the
stiffness $\gamma_{NSCHA}(T)$ equ (\ref{GAMNSCHA}) does not coincide with ${%
\widetilde J}(T) $. Chiral fluctuations exist even when $\theta^0_i=0$. 
Equ (\ref{GAMMC}) shows that $\gamma_{NSCHA}(T)$ represents the LW contribution to the stiffness constant $\gamma$, i.e the bare coupling for the phase variables. So it is natural to identify ${T\over 2\pi\gamma_{NSCHA}}$ with $T_{CG}$.
If we use $\gamma_{NSCHA}(T)$ in equ (\ref{JCG}) we find analytically 
$T_{CG}=0.198$ for the SQ lattice and $T_{CG}=0.191$ for the TR lattice, to
be compared with the MC values $0.1956$ for the SQ lattice \cite{O3} and
0.192 for the TR lattice.

To summarize our results, we have constructed a variational ensemble (NSCHA)
for fully frustrated XY systems in 2D. Testing its predictions with Monte
Carlo simulations we see that our approach yields accurate results at all
temperature -- including in the critical regime -- for quantities sensitive
to long wavelength excitations. The key ingredient of the theory is the
coupling between phase and chiral degrees of freedom and this coupling is
always relevant. In particular it causes the interaction between phase
variables to be polar-like (long range and oscillatory). As a result
renormalization schemes assuming short range couplings might not be reliable.

\noindent If a Coulomb gas temperature is introduced it appears to track
chiral variables rather than phase variables.

\noindent Monte Carlo simulations show that defects drive the transitions.
In particular chiral domains appear to affect the spinwave stiffness
constant and chiralities in a similar fashion giving support for a single
phase transition scenario.

\noindent The above results pertain to both the FFTXY and the FFSQXY
lattices suggesting universality for fully frustrated systems.

\noindent For ferromagnetic systems NSCHA still improve on SCHA. The reason
is because NSCHA incorporates fluctuations of the macroscopic phase ($%
\theta_0$) about its equibrium (zero) value, in contradistinction with SCHA.
In that sense NSCHA is a canonical ensemble as opposed to SCHA which is a
microcanonical ensemble. In that limit the Coulomb gas temperature is
associated with the bare coupling constant of the phase variables.

\section*{ACKNOWLEDGMENTS}

We enjoyed fruitful discussions on this problem with R. Eymard, E. Granato,     M. Ney-Nifle
and W.M. Saslow. Computer time on the Cray C98 was made possible by contrat
960162 from IDRIS.

\newpage


\appendix
\section{ }

In this appendix we show that in the limit of large distances $R=|\vec r_i - 
\vec r_j|$, $\widetilde J_{ij}$ given by equ (\ref{JNSCHA}) behaves as $%
\widetilde J_{ij}\sim {\frac{1}{|\vec r_i - \vec r_j|^z}} $ (equ (\ref{LRA}%
)).

We start with the NSCHA variational equation for $\widetilde J_{ij}$ :

\begin{eqnarray*}
\widetilde{J}_{ij} &=&J_{ij}\cos (\theta _i^0-\theta _j^0)e^{-\frac 12y_{ij}}
\\
&&+\frac 1{2T}\sum_{k,l}J_{ij}J_{kl}\sin (\theta _i^0-\theta _j^0)\sin
(\theta _k^0-\theta _l^0)e^{-\frac 12%
(y_{ij}+y_{kl}+y_{ik}+y_{jl}-y_{il}-y_{jk})} \\
&&+\frac 1T\sum_{k,l}J_{ik}J_{jl}\sin (\theta _i^0-\theta _k^0)\sin (\theta
_j^0-\theta _l^0)\cosh (y_{ij}+y_{kl}-y_{il}-y_{jk})e^{-\frac 12%
(y_{ik}+y_{jl})}
\end{eqnarray*}
where $y_{ij}\equiv y(\vec r_j-\vec r_i)$ is given by equ (\ref{CORRELNSCHA}%
) and where the angles $\{\theta _i^0\}$ satisfy equs (\ref{ANGSQ}, \ref
{ANGTR}).

\noindent
For the FFTXY the expression $J_{ij} \sin(\theta_{j}^{0} - \theta_{i}^{0})$
only depends upon $\vec{r}_j - \vec{r}_i$ . For the FFSQXY lattice however,
there are four different types of sites (see equ (\ref{ANGSQ})) so that $%
J_{ij} \sin(\theta_{j}^{0} - \theta_{i}^{0})$ explicitely depends upon site $%
i$. Yet, for the FFSQXY lattice the quantity $(-1)^{x_i + y_i} J_{ij}
\sin(\theta_{j}^{0} - \theta_{i}^{0})$ is independent of $i$ .

\noindent
Therefore we introduce 
\begin{equation}  \label{JBAR}
\bar{J}_{\vec{r}_{j} - \vec{r}_{i}} = w^{x_i + y_i} J_{ij}
\sin(\theta_{j}^{0} - \theta_{i}^{0}),
\end{equation}
For the triangular lattice we choose $w=1$ and for the square lattice we set 
$w=-1$. The quantity $\bar{J}$ defined in that way only depends upon $\vec{r}%
_{j} - \vec{r}_{i}$. $\bar{J}$ possesses the following symmetry properties : 
\begin{equation}  \label{JBARTR}
\bar{J}(-\vec \varepsilon) = -\bar{J}(\vec \varepsilon)
\end{equation}
for the TR lattice and 
\begin{equation}  \label{JBARSQ}
\bar{J}(-\vec \varepsilon) = \bar{J}(\vec \varepsilon),\;\;\;\;\; \bar{J}( 
\vec{u_x}) = -\bar{J}( \vec{u_y}) \\
\end{equation}
for the SQ lattice.

In the following we introduce the notations $\vec R =\vec{r}_{j} -\vec{r}%
_{i} $, $\vec {\varepsilon ^{\prime}}=\vec{r}_{l} -\vec{r}_{j}$, $\vec {%
\varepsilon}=\vec{r}_{k} -\vec{r}_{i}$. For large $R$, only the third term
in the r.h.s of equ (\ref{JNSCHA}) contributes, since $J_{ij}$ is a nearest
neighbor interaction. Using the fact that $\bar{J}_{ij}$ is independent of $%
i $, that we have the symmetry properties, equs (\ref{JBARTR},\ref{JBARSQ})
and that the value of $y_{ij}$ is the same for all the {\it nearest
neighbors $j$} of a given site i (by symmetry) we get

\begin{eqnarray*}
\widetilde{J}_{\vec R}=-{\frac 1T}e^{-y(\vec {\varepsilon _0}%
)}w^{R_x+R_y}\sum_{\vec \varepsilon ,\vec {\varepsilon ^{\prime }}}\bar J_{%
\vec \varepsilon }\bar J_{\vec {\varepsilon ^{\prime }}}\cosh \frac 12(y(%
\vec R+\vec \varepsilon +\vec {\varepsilon ^{\prime }})+y(\vec R)-y(\vec R+%
\vec \varepsilon )-y(\vec R+\vec {\varepsilon ^{\prime }}))
\end{eqnarray*}

where $\vec{\varepsilon_0}$ is any nearest neighbor vector connecting two
sites.

We denote by $\triangle y(\vec{R},\vec{\varepsilon},\vec{\varepsilon
^{\prime}})$ the following quantity:

$\triangle y(\vec R,\vec \varepsilon ,\vec {\varepsilon ^{\prime }})\equiv y(%
\vec R+\vec \varepsilon +\vec {\varepsilon ^{\prime }})+y(\vec R)-y(\vec R+%
\vec \varepsilon )-y(\vec R+\vec {\varepsilon ^{\prime }})$. In the large $R$
limit we expand $y$ in powers of $R$; with : 
\begin{eqnarray*}
\triangle _2y(\vec R,\vec \varepsilon ,\vec {\varepsilon ^{\prime }})\equiv -%
\frac 12\sum_{\alpha ,\beta }(\varepsilon _\alpha ^{\prime }\varepsilon
_\beta +\varepsilon _\alpha \varepsilon _\beta ^{\prime })\frac{\partial ^2y(%
\vec R)}{\partial R_\alpha \partial R_\beta }
\end{eqnarray*}
and 
\begin{eqnarray*}
\triangle _3y(\vec R,\vec \varepsilon ,\vec \varepsilon ^{\prime })\equiv 
\frac 16\sum_{\alpha ,\beta ,\gamma }(-\varepsilon _\alpha ^{\prime
}\varepsilon _\beta ^{\prime }\varepsilon _\gamma -\varepsilon _\alpha
^{\prime }\varepsilon _\gamma \varepsilon _\beta ^{\prime }-\varepsilon
_\beta ^{\prime }\varepsilon _\gamma ^{\prime }\varepsilon _\alpha
+\varepsilon _\alpha \varepsilon _\beta \varepsilon _\gamma ^{\prime
}+\varepsilon _\alpha \varepsilon _\gamma \varepsilon _\beta ^{\prime
}+\varepsilon _\beta \varepsilon _\gamma \varepsilon _\alpha ^{\prime })%
\frac{\partial ^3y(\vec R)}{\partial R_\alpha \partial R_\beta \partial
R_\gamma }
\end{eqnarray*}
we have that $\triangle y(\vec R,\vec \varepsilon ,\vec \varepsilon ^{\prime
})=\triangle _2y(\vec R,\vec \varepsilon ,\vec \varepsilon ^{\prime
})+\triangle _3y(\vec R,\vec \varepsilon ,\vec \varepsilon ^{\prime })+\cdot
\cdot \cdot $

\vspace {1cm} \noindent
In Appendix B we show that $\tilde{J}_{0} - \tilde{J}_{\vec q} \sim \Gamma
q^2$ for small $\vec q$ so that -- using equ (\ref{CORRELNSCHA}) -- we get $%
y(\vec{R}) \sim \log | \vec{R} | $ for large $R$. As a result

\begin{equation}
\frac{\partial^2 y(\vec{R}) }{\partial R_\alpha \partial R_\beta }\sim \frac 
1{R^2}, \\
\;\;\;\;\;\;\frac{\partial^3 y( \vec{R}) }{\partial R_\alpha \partial R_\beta \partial
R_\gamma }\sim \frac 1{R^3}
\end{equation}
Expanding $\cosh()$ we get:

\begin{equation}
\widetilde{J}_{\vec R}=-{\frac 1T}e^{-y(\vec {\varepsilon _0}%
)}w^{R_x+R_y}\sum_{\vec {\varepsilon ^{\prime }},\vec \varepsilon }\bar J_{%
\vec \varepsilon }\bar J_{\vec \varepsilon ^{\prime }}\{1+\frac 12[\frac 12%
\triangle y(\vec R,\vec \varepsilon ,\vec {\varepsilon ^{\prime }})]^2+\cdot
\cdot \cdot \}
\end{equation}
that is 

\begin{eqnarray} \label{SIGNJTILDE}
\widetilde{J}_{\vec R} &=&-{\frac 1T}e^{-y(\vec {\varepsilon _0})}
w^{R_x+R_y}\sum_{\vec {\varepsilon ^{\prime }},\vec \varepsilon }\bar J_{%
\vec \varepsilon }\bar J_{\vec {\varepsilon ^{\prime }}}\{1+\frac 18%
[\triangle _2y(\vec R,\vec \varepsilon ,\vec {\varepsilon ^{\prime }})]^2\\
&&+\frac 14[\triangle _2y(\vec R,\vec \varepsilon ,\vec {\varepsilon ^{\prime }}%
)][\triangle _3y(\vec R,\vec \varepsilon ,\vec {\varepsilon ^{\prime }})]
+\frac 18[\triangle _3y(\vec R,\vec \varepsilon ,\vec {\varepsilon
^{\prime }})]^2\}+O(R^{-7}) \nonumber
\end{eqnarray}

${\bf \bullet }$ for the triangular lattice, using equ (\ref{JBARTR}) we have

$\sum_{\vec{\varepsilon ^{\prime}},\vec{\varepsilon }} \bar{J}_{\vec{%
\varepsilon }} \bar{J}_{\vec{\varepsilon ^{\prime}}} =0$,

$\sum_{\vec{\varepsilon ^{\prime}},\vec{\varepsilon }} \bar{J}_{\vec{%
\varepsilon }} \bar{J}_{\vec{\varepsilon ^{\prime}}} \left[\triangle _2 y ( 
\vec{R},\vec{\varepsilon },\vec{\varepsilon ^{\prime}}) \right]^2=0$,

$\sum_{\vec {\varepsilon ^{\prime }},\vec \varepsilon }\bar J_{\vec 
\varepsilon }\bar J_{\vec {\varepsilon ^{\prime }}}[\triangle _2y(\vec R,%
\vec \varepsilon ,\vec {\varepsilon ^{\prime }})][\triangle _3y(\vec R,\vec 
\varepsilon ,\vec {\varepsilon ^{\prime }})]=0$.

so that

\begin{equation}
\widetilde{J}_{\vec{R}} \simeq -{\frac{1}{T}} e^{-y( \vec{\varepsilon _0})}
\sum_{\vec{\varepsilon ^{\prime}},\vec{\varepsilon }} \bar{J}_{\vec{%
\varepsilon }} \bar{J}_{\vec{\varepsilon ^{\prime}}} \frac{1}{8}
[\triangle_3 y( \vec{R},\vec{\varepsilon },\vec{\varepsilon ^{\prime}})
]^2\sim \frac{1}{R^6}
\end{equation}

${\bf \bullet}$ for the square lattice, using equ (\ref{JBARSQ}) we have

$\sum_{\vec{\varepsilon ^{\prime}},\vec{\varepsilon }} \bar{J}_{\vec{%
\varepsilon}} \bar{J}_{\vec{\varepsilon ^{\prime}}} =0$,

but $\sum_{\vec{\varepsilon ^{\prime}},\vec{\varepsilon }} \bar{J}_{\vec{%
\varepsilon}} \bar{J}_{\vec{\varepsilon ^{\prime}}} [ \triangle_2 y( \vec{R},%
\vec{\varepsilon },\vec{\varepsilon ^{\prime}} )]^2\neq 0$.

so that 
\begin{equation}
\widetilde{J}_{\vec R}\simeq -{\frac 1T}e^{-y(\vec {\varepsilon _0}%
)}(-1)^{R_x+R_y}\sum_{\vec {\varepsilon ^{\prime }},\vec \varepsilon }\bar J%
_{\vec \varepsilon }\bar J_{\vec {\varepsilon ^{\prime }}}\frac 18[\triangle
_2y(\vec R,\vec \varepsilon ,\vec {\varepsilon ^{\prime }})]^2\sim \frac 1{%
R^4}
\end{equation}

We note the sign alternation due to the $(-1)^{R_x+R_y}$ term for the FFSQXY
lattice (see Fig. 4) 

\newpage


\section{ }

In this appendix we compute $\gamma_{NSCHA}(T)$ (equ (\ref{GAMNSCHA})) and $%
\Gamma(T)$ (equ (\ref{GAMMA})), at low $T$. We show that for the FFTXY and
FFSQXY lattices $\gamma_{NSCHA}(T) - \Gamma(T) = O(T^2)$.

\vspace {1cm} \noindent
We start with equ (\ref{JNSCHA}). Using the same notations as in Appendix A
we have

\begin{eqnarray} \label{JJ}
\widetilde{J}(\vec R) &=&|J_{\vec R}|\cos \alpha ({\vec R})e^{-\frac 12y(%
\vec R)}  \\
&&-\frac 1{4T}\ \sum_{\vec r}\sum_{\vec \varepsilon ,\vec {\varepsilon
^{\prime }}}w^{r_x+r_y}\bar J_{\vec \varepsilon }\bar J_{\vec {\varepsilon
^{\prime }}}e^{-\frac 12(y(\vec \varepsilon )+y(\vec {\varepsilon ^{\prime }}%
)+y(\vec r+\vec \varepsilon +\vec {\varepsilon ^{\prime }})+y(\vec r)-y(\vec 
r+\vec \varepsilon )-y(\vec r+\vec {\varepsilon ^{\prime }}))}  \nonumber \\
&&\times (\delta (\vec R-\vec \varepsilon )+\delta (\vec R-\vec {\varepsilon
^{\prime }})+\delta (\vec R-\vec r)+\delta (\vec R-(\vec r+\vec \varepsilon +%
\vec {\varepsilon ^{\prime }}))  \nonumber \\
&&-\delta (\vec R-\vec \varepsilon -\vec r)-\delta (\vec R-\vec {\varepsilon
^{\prime }}-\vec r)) \nonumber
\end{eqnarray}

where $\delta (...)$ denotes the Kronecker delta symbol, where $\alpha ({%
\vec R})$ was defined in equs (\ref{ANGSQ},\ref{ANGTR}) and where the
expression is written in such a way as to preserve the symmetry under the
transformation $\varepsilon \leftrightarrow \varepsilon ^{\prime }$.

We now Fourier transform equ (\ref{JJ}) :

\begin{eqnarray}\label{JQ}
\widetilde{J}_{\vec q} &=&\sum_{\vec R}|J_{\vec R}|\cos \alpha ({\vec R})e^{-%
\frac 12y(\vec R)}e^{-i\vec q.\vec R}  \\
&&-\frac 1{4T}\sum_{\vec R}\sum_{\vec r}\sum_{\vec \varepsilon ,\vec {%
\varepsilon ^{\prime }}}w^{r_x+r_y}\bar J_{\vec \varepsilon }\bar J_{\vec {%
\varepsilon ^{\prime }}}e^{-\frac 12(y(\vec \varepsilon )+y(\vec {%
\varepsilon ^{\prime }})+y(\vec r+\vec \varepsilon +\vec {\varepsilon
^{\prime }})+y(\vec r)-y(\vec r+\vec \varepsilon )-y(\vec r+\vec {%
\varepsilon ^{\prime }}))}  \nonumber \\
&&\times (\delta (\vec R-\vec \varepsilon )+\delta (\vec R-\vec {\varepsilon
^{\prime }})+\delta (\vec R-\vec r)+\delta (\vec R-(\vec r+\vec \varepsilon +%
\vec {\varepsilon ^{\prime }}))  \nonumber \\
&&-\delta (\vec R-\vec \varepsilon -\vec r)-\delta (\vec R-\vec {\varepsilon
^{\prime }}-\vec r))e^{-i\vec q.\vec R} \nonumber
\end{eqnarray}
setting 
\begin{eqnarray*}
B &\equiv &\sum_{\vec R}(\delta (\vec R-\vec \varepsilon )+\delta (\vec R-%
\vec {\varepsilon ^{\prime }})+\delta (\vec R-\vec r)+\delta (\vec R-(\vec r+%
\vec \varepsilon +\vec {\varepsilon ^{\prime }})) \\
&&-\delta (\vec R-\vec \varepsilon -\vec r)-\delta (\vec R-\vec {\varepsilon
^{\prime }}-\vec r))e^{-i\vec q.\vec R}
\end{eqnarray*}
we see that $B=e^{-i\vec q.\vec \varepsilon }+e^{-i\vec q.\vec {\varepsilon
^{\prime }}}+e^{-i\vec q.\vec r}+e^{-i\vec q.(\vec r+\vec \varepsilon +\vec {%
\varepsilon ^{\prime }})}-e^{-i\vec q.(\vec r+\vec \varepsilon )}-e^{-i\vec q%
.(\vec r+\vec {\varepsilon ^{\prime }})}$

\noindent
Since we wish to compute 
\begin{equation}
\Gamma(T) = \lim_{q_x \rightarrow 0}\frac{1}{q_x^2}(\widetilde{J}(0) - 
\widetilde{J}_{\vec q. \vec{u_x}})
\end{equation}
we may expand $B$ to second order in $q$:

$B=2-(\vec q.\vec \varepsilon )(\vec q.\vec {\varepsilon ^{\prime }})+O(q^3)$

and thus :

\begin{eqnarray}\label{LIMGAM}
\Gamma (T) &=&\frac 12\sum_{\vec \varepsilon }|J_{\vec \varepsilon }|\cos
\alpha (\vec \varepsilon )(\vec {u_x}.\vec \varepsilon )^2e^{-\frac 12y(\vec 
\varepsilon )}  \\
&&+\frac 1{4T}\sum_{\vec r}\sum_{\vec \varepsilon ,\vec {\varepsilon
^{\prime }}}w^{r_x+r_y}\bar J_{\vec \varepsilon }\bar J_{\vec {\varepsilon
^{\prime }}}(\vec {u_x}.\vec \varepsilon )(\vec {u_x}.\vec {\varepsilon
^{\prime }})  \nonumber \\
&&\times e^{-\frac 12(y(\vec \varepsilon )+y(\vec {\varepsilon ^{\prime }}%
)+y(\vec r+\vec \varepsilon +\vec {\varepsilon ^{\prime }})+y(\vec r)-y(\vec 
r+\vec \varepsilon )-y(\vec r+\vec {\varepsilon ^{\prime }}))} \nonumber
\end{eqnarray}

Similarly, $\gamma_{NSCHA}(T)$ equ (\ref{GAMNSCHA}) is given by

\begin{eqnarray} \label{LIMGAMNSCHA}
\gamma _{NSCHA}(T) &=&\frac 12\sum_{\vec \varepsilon }|J_{\vec \varepsilon
}|\cos \alpha (\vec \varepsilon )(\vec \varepsilon .\vec {u_x})^2e^{-\frac 12%
y(\vec \varepsilon )}  \\
&&-\frac 1{4T}\sum_{\vec r}\sum_{\vec \varepsilon ,\vec {\varepsilon
^{\prime }}}(\vec \varepsilon .\vec {u_x})(\vec {\varepsilon ^{\prime }}.%
\vec {u_x})(|J_{\vec \varepsilon }|\cos \alpha (\vec \varepsilon )|J_{\vec {%
\varepsilon ^{\prime }}}|\cos \alpha (\vec {\varepsilon ^{\prime }}%
)-w^{r_x+r_y}\bar J_{\vec \varepsilon }\bar J_{\vec {\varepsilon ^{\prime }}%
})  \nonumber \\
&&\times e^{-\frac 12(y(\vec \varepsilon )+y(\vec {\varepsilon ^{\prime }}%
)+y(\vec r+\vec \varepsilon +\vec {\varepsilon ^{\prime }})+y(\vec r)-y(\vec 
r+\vec \varepsilon )-y(\vec r+\vec {\varepsilon ^{\prime }}))} \nonumber
\end{eqnarray}

We see that the difference between the expression of $\Gamma(T)$ and the
expression of $\gamma_{NSCHA}(T)$ comes from the term proportionnal to $\cos
\alpha(\vec \varepsilon) \cos \alpha(\vec{\varepsilon ^{\prime}})$.

With the notations:

\begin{eqnarray*}
A &=&\frac 12\sum_{\vec \varepsilon }|J_{\vec \varepsilon }|\cos (\alpha (%
\vec \varepsilon ))(\vec u_x\vec \varepsilon )^2e^{-\frac 12y(\vec 
\varepsilon )} \\
S &=&\frac 14\sum_{\vec r}\sum_{\vec \varepsilon ,\vec \varepsilon ^{\prime
}}w^{r_x+r_y}\bar J_{\vec \varepsilon }\bar J_{\vec \varepsilon ^{\prime }}(%
\vec u_x.\vec \varepsilon )(\vec u_x.\vec \varepsilon ^{\prime }) \\
&&\times e^{-\frac 12(y(\vec \varepsilon )+y(\vec \varepsilon ^{\prime })+y(%
\vec r+\vec \varepsilon +\vec \varepsilon ^{\prime })+y(\vec r)-y(\vec r+%
\vec \varepsilon )-y(\vec r+\vec \varepsilon ^{\prime }))} \\
C &=&\frac 14\sum_{\vec r}\sum_{\vec \varepsilon ,\vec \varepsilon ^{\prime
}}|J_{\vec \varepsilon }|\cos (\alpha (\vec \varepsilon ))|J_{\vec {%
\varepsilon ^{\prime }}}|\cos (\alpha (\vec {\varepsilon ^{\prime }}))(\vec u%
_x.\vec \varepsilon )(\vec u_x.\vec \varepsilon ^{\prime }) \\
&&\times e^{-\frac 12(y(\vec \varepsilon )+y(\vec \varepsilon ^{\prime })+y(%
\vec r+\vec \varepsilon +\vec \varepsilon ^{\prime })+y(\vec r)-y(\vec r+%
\vec \varepsilon )-y(\vec r+\vec \varepsilon ^{\prime }))}
\end{eqnarray*}

$\gamma _{NSCHA}(T)$ and $\Gamma(T)$ read :

\begin{eqnarray*}
\gamma_{NSCHA}(T) &=& A-\frac{1}{T}C+\frac{1}{T}S  \nonumber \\
\Gamma(T) &=& A+\frac{1}{T}S
\end{eqnarray*}

Expanding $A$, $C$ and $S$ in $T$ yields:

$C=C^0+C^1T+C^2T^2+O(T^3)$ and $S=S^0+S^1T+S^2T^2+O(T^3)$

We set $g(\vec r) \equiv y(\vec r)/T$ such that $g(\vec r) $ approaches a
finite limit as $T\rightarrow 0$. \vspace {0.5cm}

\underline{to order $T^0$ :}

\begin{eqnarray*}
C^0 &=&\frac 14\sum_{\vec r}\sum_{\vec \varepsilon ,\vec \varepsilon
^{\prime }}|J_{\vec \varepsilon}| \cos \alpha(\vec \varepsilon) |J_{\vec{%
\varepsilon ^{\prime}}}| \cos \alpha(\vec{\varepsilon^{\prime}}) (\vec 
\varepsilon .\vec u_x)(\vec \varepsilon ^{\prime }.\vec u_x) =0  \nonumber \\
S^0 &=&-\frac 14\sum_{\vec r}\sum_{\vec \varepsilon ,\vec \varepsilon
^{\prime }}w^{r_x + r_y} \bar{J}_{\vec \varepsilon } \bar{J}_{\vec 
\varepsilon ^{\prime }} (\vec \varepsilon .\vec u_x)(\vec \varepsilon
^{\prime }.\vec u_x) =0
\end{eqnarray*}

$C^0$ is clearly zero when one sums over $\vec \varepsilon $ and $\vec 
\varepsilon ^{\prime }$

\noindent
$S^0=0$ for the SQ lattice because $w=-1$ so that $\sum_{\vec r}w^{r_x +
r_y} = 0$; $S^0=0$ for the TR lattice because $\sum_{\vec \varepsilon} \bar{J%
}_{\vec \varepsilon}(\vec{\varepsilon}.\vec{u_x}) = 0$ \vspace {0.5cm}

\underline{to order $T^1$ :}

\begin{eqnarray*}
C^1 &=&\frac 14\sum_{\vec r}\sum_{\vec \varepsilon ,\vec \varepsilon
^{\prime }}|J_{\vec \varepsilon}| \cos \alpha(\vec \varepsilon) |J_{\vec 
\varepsilon}| \cos \alpha(\vec \varepsilon) (\vec \varepsilon .\vec u_x)(%
\vec \varepsilon ^{\prime }.\vec u_x)  \nonumber \\
& &.\frac 12(g(\vec \varepsilon )+g(\vec \varepsilon ^{\prime })+g(\vec r+%
\vec \varepsilon +\vec \varepsilon ^{\prime })+g(\vec r)-g(\vec r+\vec 
\varepsilon )-g(\vec r+\vec \varepsilon ^{\prime })))  \nonumber \\
&=&0
\end{eqnarray*}

\begin{eqnarray*}
S^1 &=&\frac 14\sum_{\vec r}\sum_{\vec \varepsilon ,\vec \varepsilon
^{\prime }}w^{r_x + r_y} \bar{J}_{\vec \varepsilon } \bar{J}_{\vec 
\varepsilon ^{\prime }} (\vec \varepsilon .\vec u_x)(\vec \varepsilon
^{\prime }.\vec u_x)  \nonumber \\
&&\ (-\frac 12)(g(\vec \varepsilon )+g(\vec \varepsilon ^{\prime })+g(\vec r+%
\vec \varepsilon +\vec \varepsilon ^{\prime })+g(\vec r)-g(\vec r+\vec 
\varepsilon )-g(\vec r+\vec \varepsilon ^{\prime }))  \nonumber \\
&=&0
\end{eqnarray*}

Both $C^1$ and $S^1$ equal zero, owing to the parity in $\vec \varepsilon $
and $\vec \varepsilon ^{\prime }$, and using the fact that $\sum_{\vec r}g(%
\vec r+\vec a)=\sum_{\vec r}g(\vec r)$ whenever $\vec a$ is any vector
connecting sites of the lattice. \vspace {0.5cm}

\underline{to order $T^2$ :}

\begin{eqnarray*}
C^2 &=&\frac 14\sum_{\vec r}\sum_{\vec \varepsilon ,\vec \varepsilon
^{\prime }}|J_{\vec \varepsilon}| \cos \alpha(\vec \varepsilon) |J_{\vec{%
\varepsilon ^{\prime}}}| \cos \alpha(\vec{\varepsilon ^{\prime}}) (\vec 
\varepsilon .\vec u_x)(\vec \varepsilon ^{\prime }.\vec u_x)  \nonumber \\
& &.\frac 18 (g(\vec \varepsilon )+g(\vec \varepsilon ^{\prime })+g(\vec r+%
\vec \varepsilon +\vec \varepsilon ^{\prime })+g(\vec r)-g(\vec r+\vec 
\varepsilon )-g(\vec r+\vec \varepsilon ^{\prime }))^2
\end{eqnarray*}

\begin{eqnarray*}
S^2 &=&\frac 14\sum_{\vec r}\sum_{\vec \varepsilon ,\vec \varepsilon
^{\prime }}w^{r_x + r_y} \overline{J}_{\vec \varepsilon } \overline{J}_{\vec 
\varepsilon ^{\prime }}  \nonumber \\
& &.\frac 18(g(\vec \varepsilon )+g(\vec \varepsilon ^{\prime })+g(\vec r+%
\vec \varepsilon +\vec \varepsilon ^{\prime })+g(\vec r)-g(\vec r+\vec 
\varepsilon )-g(\vec r+\vec \varepsilon ^{\prime }))^2
\end{eqnarray*}
Using the same properties as for the terms of order $T$ we find:

\begin{eqnarray*}
C^2 &=&0  \nonumber \\
S^2 &=&\frac 14\sum_{\vec r}\sum_{\vec \varepsilon ,\vec \varepsilon
^{\prime }}w^{r_x + r_y} \bar{J}_{\vec \varepsilon } \bar{J}_{\vec 
\varepsilon ^{\prime }}  \nonumber \\
& &.\frac 12 g(\vec r)g(\vec r+\vec \varepsilon +\vec \varepsilon ^{\prime })
\end{eqnarray*}

Also expanding $A$ to order $T$ gives :

\[
A=\frac 12\sum_{\vec \varepsilon }|J_{\vec \varepsilon}| \cos \alpha(\vec 
\varepsilon).(\vec u_x\vec \varepsilon )^2 (1-\frac T2g(\vec \varepsilon ))
+ O(T^2) 
\]

From these calculations we deduce that $\gamma_{NSCHA}(T)= \Gamma(T)$ to
order $T$ ( $C=O(T^3)$ ). At this order we may simply replace $\widetilde{J}%
_{\vec q}$ by ${J}_{\vec q}$ in the expression of $g(\vec r)$ and we find
that (equ (\ref{RIGID}))

\begin{equation}
\gamma_{NSCHA}(T)=\gamma_0(1-{\frac{T}{T_{c0}}})
\end{equation}

with

${\bf \bullet}$ For the triangular lattice $\gamma_0=\sqrt 3/2 J$ and $%
T_{c0}={\frac{1}{ 4/3-{\frac{3\sqrt 3}{2\pi}}}}J\sim 1.975J$

${\bf \bullet}$ For the square lattice $\gamma_0=\sqrt 2/2J$ and $T_{c0}= {%
\frac{1}{{\frac{\sqrt 2}{2}}-{\frac{\sqrt 2}{2\pi}}}}J\sim 2.075J$



\end{document}